\documentclass[epj]{webofc}
\usepackage[varg]{txfonts}   
\usepackage{epsfig}
\woctitle{MENU 2013}
\begin{document}
\title{Improved description of the $\pi N$-scattering
phenomenology in \\ covariant baryon chiral
perturbation theory}
%
%

\author{Jose Manuel Alarc\'on\inst{1}\fnsep\thanks{\email{alarcon@kph.uni-mainz.de}}
}

\institute{Cluster of Excellence PRISMA, Institut f\"ur Kernphysik, \\Johannes Gutenberg-Universit\"at, Mainz D-55099, Germany 
          }

\abstract{%
  We highlight some of the recent advances in the application of chiral effective field theory (chiral EFT) with baryons to the $\pi N$ scattering process.
  We recall some problems that cast doubt on the applicability of chiral EFT to $\pi N$ and show how the relativistic formalism, once the $\Delta(1232)$-resonance is included as an explicit degree of freedom, solves these issues. 
  Finally it is shown how this approach can be used to extract the $\sigma$-terms from phenomenological information.
}
\maketitle
\section{Introduction}
\label{intro}

Our knowledge of nuclear interactions is involved, to a greater or less extent, in many of the current experimental tests of the fundamental interactions as well as in searches for new physics.
Unfortunately, the fundamental theory describing these interactions (QCD) becomes strongly coupled in the regime of energies where nuclear physics is interested in.
However, one can construct an effective field theory embodying the relevant symmetries of the fundamental theory for the range of energies of interest. 
In this way, one can work in a quantum field theory (QFT) that has the same physical content as the fundamental theory in that range of energies.
For QCD, this was done by Gasser and Leutwyler in \cite{Gasser:1983yg} for the Goldstone sector  of the spontaneously broken chiral symmetry (the low energy symmetry of QCD), while the extension to the one-nucleon sector was done in \cite{Gasser:1987rb} by Gasser, Sainio and Svarc (GSS). 
In this reference, it was pointed out the difficulty to keep the perturbative treatment (the main advantage of the EFT) together with the presence of a heavy scale (i. e. the nucleon mass).
After some years, it was shown that the terms breaking the power counting (due to that heavy scale) have no physical meaning (are just analytical pieces that redefine the bare parameters of the original Lagrangian), while the unitary corrections respect the counting in the soft scale \cite{Gegelia:1999gf}.
This solution to the power counting problem of chiral EFT with baryons allows us to work in a framework where all the good analytical properties of the original relativistic QFT formulation of GSS are preserved. 
However, this is not enough to ensure the convergence of the perturbative expansion.
A required feature of EFT is to have a proper separation of scales so that the non-perturbative effects can be integrated out. 
In the baryonic sector this is another complication, since the mass gap between the nucleon and the $\Delta(1232)$ (to which the $\pi N$ system couples strongly) is quite small compared to the chiral symmetry breaking scale. 
This reduces considerably the range of applicability of an EFT with only $\pi$ and $N$ as degrees of freedom, since the $\Delta(1232)$ peak is very close to the $\pi N$ threshold.
One can avoid these limitations by including this resonance as a dynamical degree of freedom, which introduces a new extra scale to the theory dictated precisely by this mass difference ($\delta\equiv m_\Delta - m_N$, being $m_\Delta$ and $m_N$ the masses of the $\Delta(1232)$ and nucleon).
This combination of the relativistic formalism with the inclusion of the $\Delta(1232)$ has been very useful to understand some of the fundamental nuclear processes \cite{Lensky:2009uv,Alarcon:2012kn} and the nuclear structure \cite{Lensky:2009uv,Alarcon:2011zs,Alarcon:2012nr,Bernard:2012hb} on chiral symmetry grounds.
Further improvements in the chiral EFT description of the fundamental nuclear reactions like Compton scattering, pion photoproduction  and $\pi N$ scattering will help to understand the nuclear structure in terms of the chiral symmetry, as well as to improve the chiral description of the $NN$ and many-nucleons interactions. 
Any progress in this direction will be positive for {\it ab initio} calculations of more complicated nuclear processes by means, for example, of nuclear lattice EFT \cite{Epelbaum:2009zsa}.

\section{Chiral EFT with baryons: Theoretical aspects}
\label{Sec:theo}

The aim of perturbation theory is to calculate small corrections to a solvable problem. 
For the par\-ti\-cu\-lar case of $\pi N$ scattering, one calculates corrections to the soft limit, in which the external momenta and quark masses are zero. 
In that limit, the meson cloud decouples from the nucleon leading to a trivial $T$-matrix. 
Using as kinematical variables $\nu\equiv (s-u)/4m_N$ and $t$, the soft limit would correspond to the point $(\nu=0, t=0)$ in the Mandelstam plane.
Therefore, the perturbative (low energy) expansion of the amplitudes is performed around this point.
For $\pi N$, the scattering amplitude can be written in terms of four functions: $D(\nu,t)^\pm$, $B(\nu,t)^\pm$ (see \cite{Alarcon:2012kn} for definitions).
Once the Born terms are subtracted, these functions allow a low energy expansion around the soft point that can be written in general form as

\begin{equation}
X^{\pm}(\nu,t)=x_{00}^{\pm}+x_{10}^{\pm}\nu^2+ x_{01}^{\pm}t+ x_{20}^{\pm}\nu^4+  x_{02}^{\pm}t^2+\ldots, \label{Eq:subthrExp} 
\end{equation}

with $X^\pm=\bar{D}^+,\bar{D}^-/\nu,\bar{B}^+/\nu,\,\bar{B}^-$, the Born-subtracted scattering amplitudes. 
The coefficients of the low-energy chiral expansion can be in turn related to those of Eq.~\eqref{Eq:subthrExp}, the so-called subthreshold coefficients.

One of the most serious problems encountered by the Lorentz invariant formulation of baryon chiral EFT proposed in Ref.~\cite{Becher:1999he}, called Infrared Regularization, was the disagreement in their extraction of the subthreshold coefficients \cite{Becher:2001hv} respect to the dispersive calculation of Ref.~\cite{KA85}. 
Since the determination of Ref.~\cite{Becher:2001hv} used information in the physical region, this would mean that chiral EFT was not able to connect the soft point to the physical region, at least up to one loop accuracy (the accuracy of the calculation).
This led the authors to doubt about the reliability of chiral EFT with baryons applied to $\pi N$ scattering, since the chiral expansion is performed around the soft point.

However, we showed in \cite{Alarcon:2012kn} that this problem was mainly due to the omission of the $\Delta(1232)$ in the theoretical approach.
This is because of the strong coupling at low energies of this resonance to the $\pi N$ system, what prevents a decoupled description in which the $\pi$ and $N$ are the only degrees of freedom and the $\Delta(1232)$ is integrated out.
Also, we show in the same reference that keeping the good analytical properties of the relativistic formulation is important in order to extract correctly the $\pi N$ scattering phenomenology.
Once the $\Delta(1232)$ is included as a degree of freedom in the relativistic formalism, the chiral expansion shows a much better convergence. 
This improved convergence allowed us to extract, for first time in the literature, values for the subthreshold coefficients that are in general in good agreement with the partial wave (PW) solutions \cite{Alarcon:2012kn}, proving that baryon chiral EFT is reliable to study the $\pi N$ scattering process at low energies.

\section{Applications}
\label{Sec:Applications}

It is important to highlight some of the practical benefits of the improved converge achieved in this approach.
Perhaps one of the most interesting quantities to study from $\pi N$ scattering is the pion-nucleon sigma term, $\sigma_{\pi N}\equiv \hat{m}\langle N | (\bar{u}u + \bar{d}d)   | N \rangle/2m_N$ (with $\hat{m}\equiv (m_u+m_d)/2$ the light quarks average and $m_N$ the nucleon mass).
This gives information about the scalar coupling of the nucleon and is an important quantity for hadronic physics as well as for direct detection of dark matter \cite{DMDetection}.
Using this approach we showed in \cite{Alarcon:2011zs} that the modern $\pi N$ partial wave analyses (PWAs) \cite{WI08, EM06} point to a relatively large value of the sigma term,
\begin{align}
\sigma_{\pi N}=59(7)~\text{MeV}.
\end{align}
It was also shown that this value is favoured by updated $\pi N$ phenomenology, like the recent extraction of the scalar-isoscalar scattering length $a_{0\, +}^{+}$ obtained in \cite{Baru:2010xn}, which is closely related to the value of $\sigma_{\pi N}$ (see Ref.~\cite{Gasser:1990ce,Pavan:2001wz}).

\begin{figure}
\begin{center}
 \epsfig{file=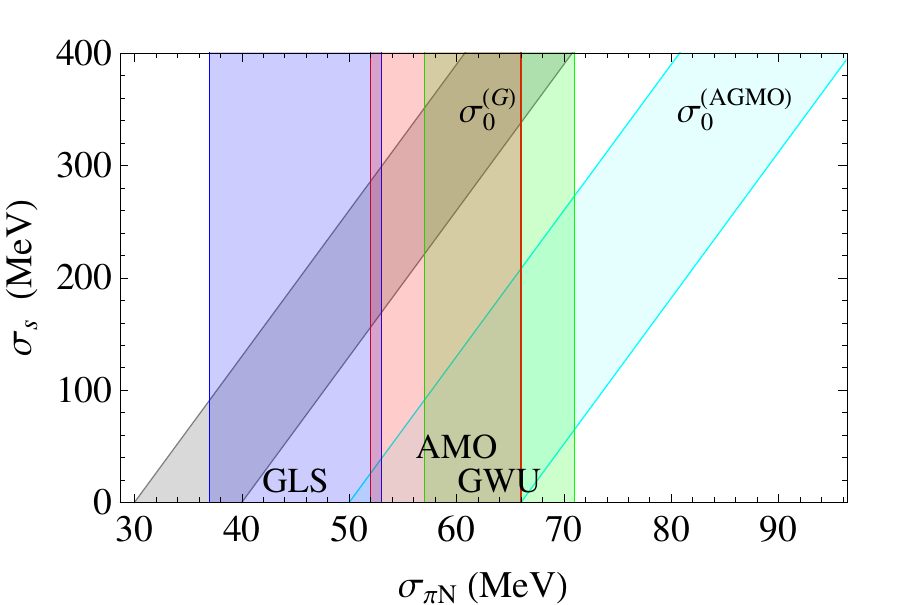,width=.63\textwidth,angle=0} 
\caption[pilf]{\small Some of the commonly used values for the sigma terms. $\sigma_{\pi N}(\text{GLS})=45(8)$~MeV~\cite{Gasser:1990ce} (blue band), $\sigma_{\pi N}(\text{AMO})=59(7)$~MeV~\cite{Alarcon:2011zs} (red band), $\sigma_{\pi N}(\text{GWU})=64(7)$~MeV~\cite{Pavan:2001wz} (green band), $\sigma_{s}^{(\text{G})}=35(5)$~MeV~\cite{Gasser:1980sb} (grey band), $\sigma_{s}^{(\text{AGMO})}=58(7)$~MeV~\cite{Alarcon:2012nr} (light blue band).\label{Fig:sigma-plane}} 
\end{center}
\end{figure}

On the other hand, given a value for $\sigma_{\pi N}$, one can infer the value of $\sigma_s\equiv m_s \langle N | \bar{s}s | N \rangle /2m_N$ through the $SU(3)_F$ breaking of the baryon octet masses \cite{Alarcon:2012nr}. 
Since the breaking in this sector is sizeable, we computed it using the relativistic formalism with the explicit inclusion of the decuplet in order to improve the convergence.
In this way, we obtained \cite{Alarcon:2012nr}
\begin{align}
\sigma_{s}=16(80)(60)~\text{MeV},\hspace{2cm} y \equiv \frac{2\langle N | \bar{s}s | N \rangle}{\langle N | (\bar{u}u+\bar{d}d) | N \rangle}=0.02(13)(10),
\end{align}
where the first error is statistical and the second one systematic (estimation of $\mathcal{O}(p^4)$ corrections in the $SU(3)_F$ breaking calculation).
These results are summarized in Fig.~\ref{Fig:sigma-plane}, together with some of the commonly used values for the sigma terms.
The vertical bands show the phenomenological extractions of $\sigma_{\pi N}$ reported in \cite{Gasser:1990ce} (GLS), \cite{Alarcon:2011zs} (AMO), and \cite{Pavan:2001wz} (GWU).
The diagonal lines show, on the other hand, the determinations of $\sigma_0\equiv \hat{m}\langle N|  \bar{u}u+\bar{d}d - 2 \bar{s}s |N \rangle/2 m_N$ of references \cite{Gasser:1980sb} ($\sigma_0^{(G)}$) and \cite{Alarcon:2012nr} ($\sigma_0^{(AGMO)}$), both based on the octet masses splitting.
Another commonly used value is the Heavy Baryon result $\sigma_0=36(7)$~MeV published in Ref.~\cite{Borasoy:1996bx}, but since numerically is very close to $\sigma_0^{(G)}$, we will use the latter for practical comparisons.
Fig.~\ref{Fig:sigma-plane} allows to analyze easily the situation for $\sigma_s$ with the different combinations of $\sigma_{\pi N}$ and $\sigma_{0}$. 
It is shown that the modern determinations of $\sigma_{\pi N}$ and $\sigma_{0}$ (AMO, GWU and $\sigma_{0}^{(AGMO)}$) point to to a small $\sigma_s$, while the older ones (GLS and $\sigma_{0}^{(G)}$) tend to give a slightly larger strangeness content.

\section{Summary and Conclusions}
\label{Sec:Summary}

Chiral EFT with baryons is an excellent tool to investigate the fundamental nuclear reactions at low energy as well as the low energy structure of the nucleon.
Here we highlighted some of the recent developments of the theory in the $\pi N$ sector, one of the fundamental reactions needed as input in more complex nuclear reactions as $NN$ scattering and many-nucleon interactions.
These developments allow a more reliable extraction of the LECs, which encode the information of the low energy $\pi N$ interaction.
These give predictions of the phenomenology in good agreement with independent determinations \cite{Alarcon:2012kn}.
Improvements in the baryonic sector of chiral EFT are interesting, among other things, to unveil the low energy structure of the nucleon \cite{Lensky:2009uv,Alarcon:2011zs,Alarcon:2012nr,Bernard:2012hb} or for theoretical calculations where the hadronic interaction of the nucleon is relevant at low energies. 
A good example of the latter is the usefulness of baryon chiral EFT in the {\it prediction} of the contribution of the proton polarizabilities to the muonic hydrogen Lamb shift \cite{Nevado:2007dd,Alarcon:2013cba}.
Therefore we consider it is worthy to apply the same approach to other fundamental nuclear processes where it has not been considered yet.

\end{document}